\documentstyle[multicol,osa,epsfig]{revtex}
\title{Low work function of the (1000) Ca$_2$N surface}
\author{M.A. Uijttewaal, G.A. de Wijs and R.A. de Groot\\
\emph{NSRIM, Toernooiveld 1, 6525 ED, Nijmegen, The Netherlands}}
\begin{document}
\maketitle
\newcommand{\E}[1]{$E_{\mathrm{ #1 }}$}

\begin{abstract} 
Polymer diodes require cathodes that do not corrode the polymer but do have low work function to minimize the electron injection barrier. First-principles calculations demonstrate that the work function of the (1000) surface of the compound Ca$_2$N is half an eV lower than that of the elemental metal Ca (2.35 vs.\ 2.87~eV). Moreover its reactivity is expected to be smaller. This makes Ca$_2$N an interesting candidate to replace calcium as cathode material for polymer light emitting diode devices.  [PACS: 73.30 +y]
\end{abstract}

\begin{multicols}{2}
One of the great challenges for polymer light emitting diodes is the electron injection barrier as well as the performance degradation caused by chemical reactions of the cathode with the polymer.\cite{chem} State-of-the-art devices\cite{stateart} use PPV (poly phenylene vinylene) as electro luminescence material. Often the cathode is made out of calcium, because of its low work function (2.87~eV)\cite{WFCa} and the presumed alignment of its Fermi level with the lowest unoccupied molecular orbital of PPV (electron affinity 2.73-2.8~eV).\cite{elafPPV1,elafPPV2} Biases actually used are much higher than the optical band gap of PPV (2.4~eV)\cite{EgPPV} namely in the order of 10~V,\cite{chem} which suggests that there is some barrier formation between the cathode and the polymer. This is further confirmed by the observation that the decrease of performance is much less for devices in which the cathode is formed at a small residual oxygen pressure.\cite{oxygen} In such a device some oxidation prevents reaction of the cathode with the polymer.

It would be a major breakthrough when stable, low work function metals are found. Sub-nitrides are fascinating candidates because, as in the case of cesium sub-oxides,\cite{CsO,b} it is expected that through quantum confinement their work functions are lower than those of the elemental metals while at the same time ionic bonding reduces reactivity.

The Ca$_2$N crystal\cite{Red} and electronic structure (bulk and single slab)\cite{Gil} are known. It is an ionic compound, built out of slabs of alternating (Ca-N-Ca) hexagonal layers. The interslab distance between Ca layers is 3.81~\AA\ and Ca layers inside a slab are only 2.45~\AA\ apart. This is much smaller than the ordinary fcc Ca (111) layer distance (3.18~\AA) and can be attributed to the ionic binding. The band structure shows fully occupied N 2s and N 2p states and a quasi 2D free-electron state in the space between the slabs which further confirms the idea that this is an ionic compound.

In this paper the Ca$_2$N work function ($\Phi$) and surface energy (\E{S}) are determined for both the (0001) and the (1000) surface. An indication of the materials stability is obtained from its binding energy and degree of surface relaxation. 
\vspace{-0.4cm}
\begin{figure}
\centerline{\includegraphics[angle=-90, width=9cm]{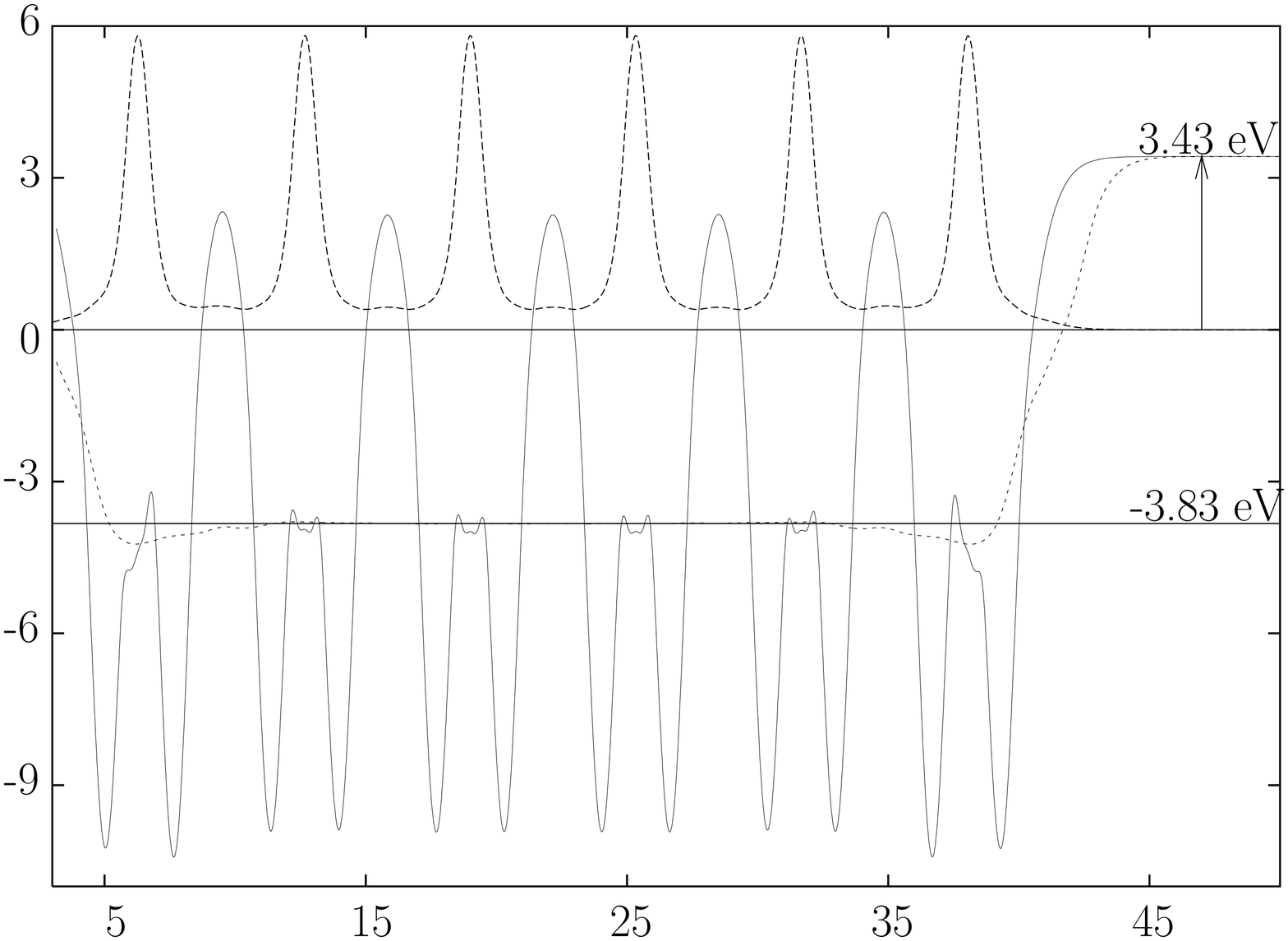}}
\caption{Charge density (dashed line, arb. units) averaged over (0001) planes and potentials (relative to $E_{\text{F}}$, eV) averaged over (0001) planes and bulk unit cells (solid/dotted lines respectively) as function of position (\AA). The horizontal line at -3.83~eV refers to the average bulk potential.}
\label{WF1}
\end{figure}
\vspace{-0.2cm}
First-principles calculations were carried out based on density functional theory in the local density approximation\cite{DFT,DFT2} with generalized gradient corrections.\cite{GGA} The total energy and molecular dynamics program VASP (Vienna \emph{Ab-initio} Simulation Package)\cite{VASP,VASP2} was used which has the projector augmented wave method\cite{PAW,PAW2} implemented. Non-linear core corrections\cite{core} were applied for calcium. The unit cells contained 18 and 42 atoms for the (0001) and the (1000) surfaces respectively. The Kohn-Sham orbitals were expanded in plane waves with cutoffs of 37~Ry. $12\cdot12\cdot1$ and $1\cdot 12\cdot 4$ Monkhorst-Pack\cite{Monk} k-point grids respectively were used to sample the Brillouin zones resulting in 74 and 26 k-points respectively in their irreducible parts.

$\Phi$ is defined as the minimum amount of energy it costs to extract an electron from a metal. The fact that it depends on the type of surface, is somewhat counter intuitive, it seems to contradict the conservation of energy. However, the differences in work function are compensated by differences in kinetic energy of the electron at large distances.\cite{Fall}
\vspace{-0.2cm}
\begin{figure}
\centerline{\includegraphics[angle=-90,width=9cm]{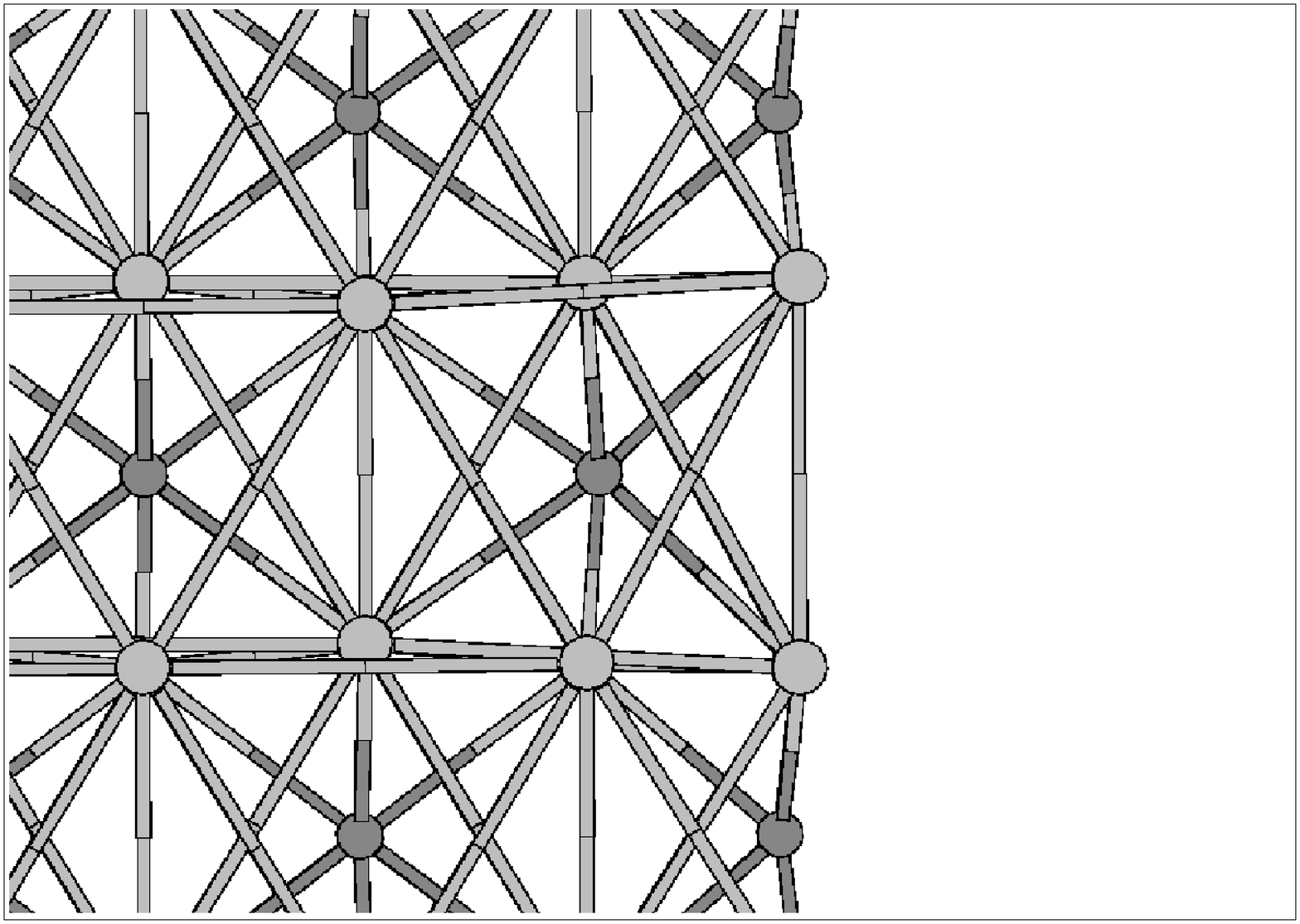}}
\caption{Atomic positions after relaxation of the (1000) surface as seen from the (1206) direction. Dark circles are nitrogens and light ones calcium.}
\label{pos3}
\end{figure}
\vspace{-0.2cm}
In order to determine $\Phi$, two numbers are needed: the maximum energy of an electron inside the material, defined to be the Fermi level ($E_{\text{F}}$), and the minimum energy outside the material, the vacuum potential. Because in a finite slab calculation $E_{\text{F}}$ cannot be accurately determined, this number is first obtained from a bulk calculation. Then for the finite slab the average potential of the middle layers (which \emph{is} an accurate quantity) must be aligned with that of the bulk.\cite{acc} 
In this way the work function for the (0001) surface is determined using 6 slabs of Ca-N-Ca and 19~\AA\ of vacuum.\cite{per} It can be extracted from Fig.~\ref{WF1} and equals 3.43~eV.

\begin{figure}
\centerline{\includegraphics[angle=-90, width=9cm]{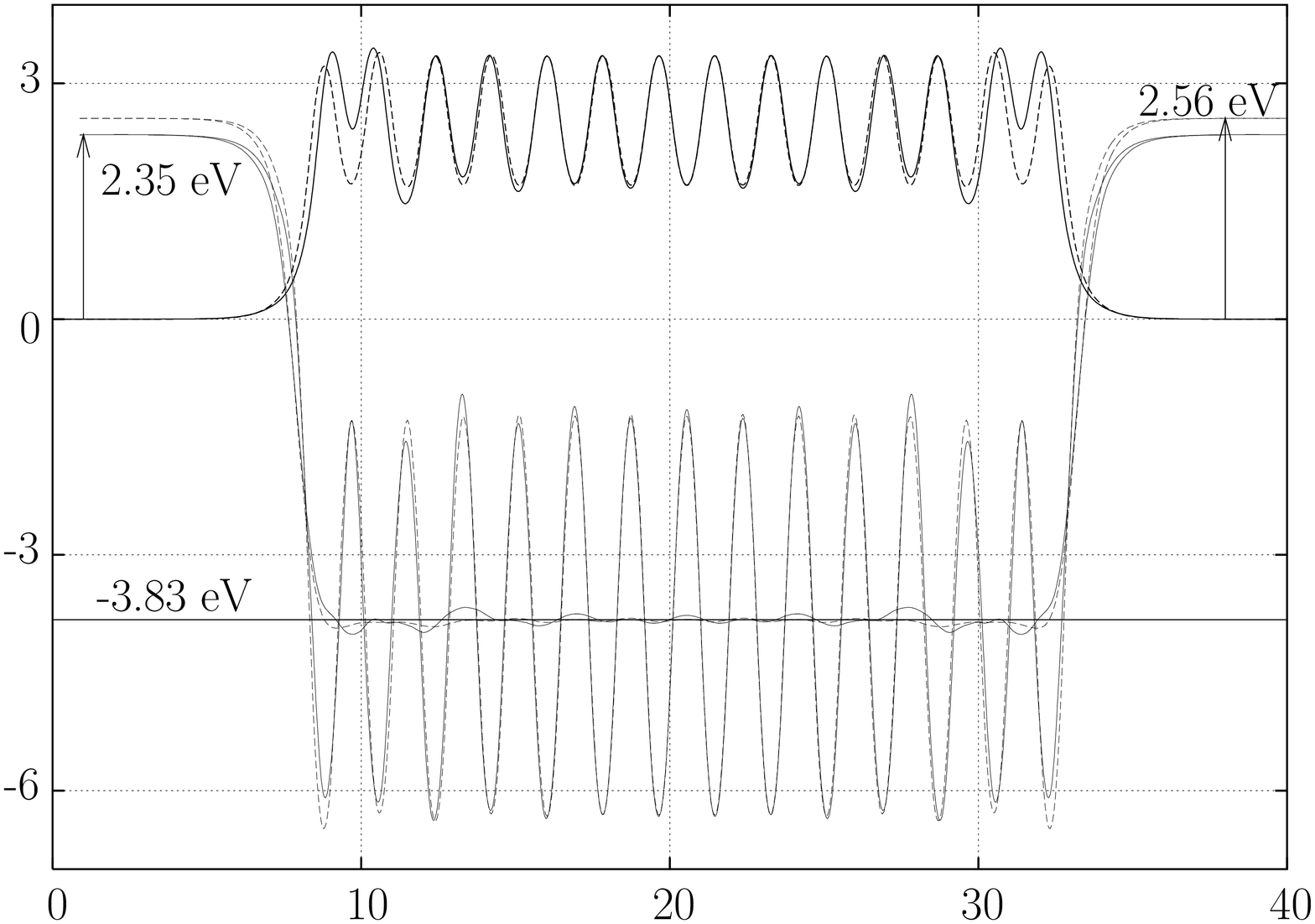}}
\caption{Charge densities averaged over (1000) planes (arb. units) and potentials (relative to $E_{\text{F}}$, eV) averaged over (1000) planes and bulk unit cells, for both the relaxed (solid lines) and the unrelaxed (dashed lines) structure as function of position (\AA). The potentials converge to 2.35~eV and 2.56~eV in the vacuum respectively. The horizontal line is the average bulk potential.}
\label{WF2}
\end{figure}

\begin{figure}
\centerline{\includegraphics[angle=-90, width=9cm]{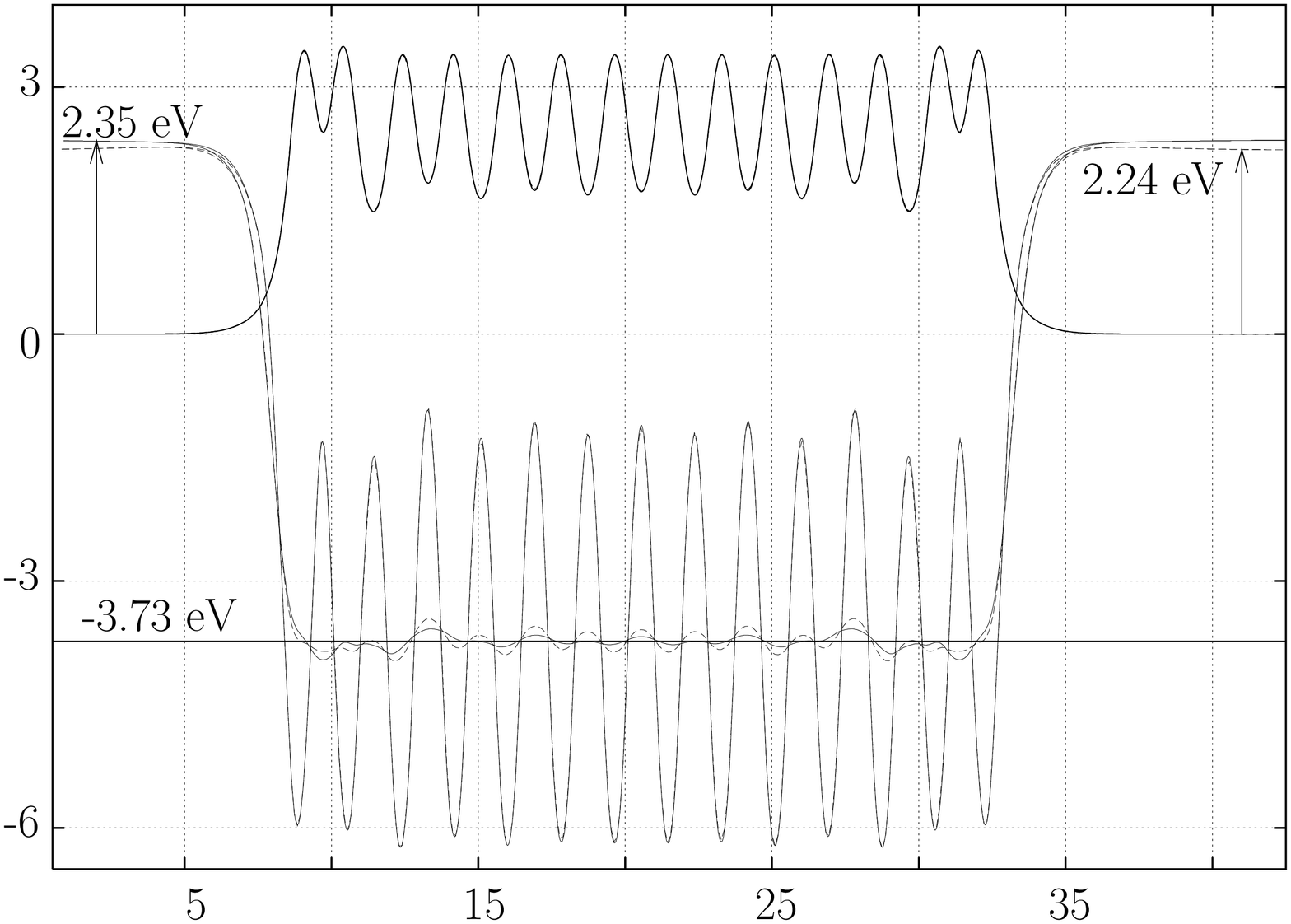}}
\caption{Charge densities averaged over (1000) planes (arb. units) and potentials (relative to $E_{\text{F}}$, eV) averaged over (1000) planes and bulk unit cells, for both the relaxed (solid lines) and the unrelaxed (dashed lines) structure for a 2\% larger $c$ axis as function of position (\AA). The potentials converge to 2.35~eV and 2.24~eV in the vacuum respectively. The horizontal line is the average bulk potential and it can be seen that relaxation has no observable effect on the charge density.}
\label{WF3}
\end{figure}

Another unit cell was made to find $\Phi$ for the (1000) surface of Ca$_2$N. It contained 14 layers and 20~\AA\ of vacuum. At this surface several bonds are severed and it is necessary to relax the atomic positions. The result is shown in Fig.~\ref{pos3} where the view is along the displacement vector from one slab to the next. The nitrogen atoms at the surface are displaced a little (0.24~\AA) inwards, the nitrogens in the second layer a little (0.17~\AA) outwards, while the surface calciums tend to move in the direction of the surface nitrogens. This can be understood from an ionic point of view. Nitrogen favors an environment of high electron density while calcium shows the opposite trend. Although the structure has changed very little, the effect of the relaxation on $\Phi$ is substantial (Fig.~\ref{WF2}). Before relaxation we find a work function of 2.56~eV, already lower than that of pure calcium, but after relaxation it has even decreased to 2.35~eV. Not only is this comparable to the smallest work function for an element (Cs, 2.14~eV),\cite{WFCa} but also the difference of 1.08~eV with the (0001) surface is larger than that between any two surfaces of tungsten,\cite{WFW} which is known for its large work function anisotropy. Both the drop in $\Phi$ after relaxation and the large surface-anisotropy can be explained by the Smoluchowski model\cite{Smol} and \emph{not} by quantum confinement as this is a bulk model that does not say anything about anisotropies and/or relaxation. According to Smoluchowski on the other hand, the more ``open'' a surface is the more the charge density can be smoothed, and this lowers $\Phi$. However more-open surfaces tend to have higher surface energies\cite{surf} and this makes them less stable. For the (0001) and (relaxed) (1000) surface we find \E{S} is 2.7 and 5.3 eV/nm$^2$ respectively which confirms the trend.
\vspace{-0.2cm}
\vspace{-0.2cm}
Because in Ref.~\onlinecite{Red} it is suggested Ca$_2$N can also be made with a larger $c$ axis (interslab separation), we examined whether this changes the 
(1000) work function. After enlarging $c$ with 2\% and subsequent relaxation of 
the atoms, the structural changes are small and the trends already indicated are a little more pronounced. The slab thickness however has decreased to almost 
(+0.5\%) its former value. The averaged charge density and local potential are 
plotted in Fig.~\ref{WF3}. Before relaxation $\Phi$ is decreased from 2.35 to 2.24~eV, but relaxation increases it again to 2.35~eV. Again this is in
accordance with the Smoluchowski trend because a thicker slab makes the surface more open. We conclude that the work function does not (significantly) change when $c$ is enlarged. 
\vspace{-0.2cm}
\vspace{-0.2cm}
Finally the reactivity of Ca$_2$N is briefly discussed. Inspection of the binding energy shows that it is 5.6~eV per formula unit (with respect to half a nitrogen molecule and 2 calcium atoms) for a single slab and 0.6~eV more for the bulk. This gives evidence that the reactivity of Ca$_2$N is smaller than that of calcium, of which the binding energy is just 1.9~eV per atom. It was shown earlier that relaxation of the atomic positions after bonds are cut or $c$ is enlarged, does not have a substantial effect on the structure. These facts taken together, give a indication that the reactivity of this sub-nitride is less than that of pure calcium, although it is still air sensitive.\cite{Red}

In summary, using first-principles calculations, we have shown that the work 
function of Ca$_2$N has an anisotropy of more than 1~eV, with a minimum of 2.35~eV (cf.\ Ca, $\Phi$ = 2.87~eV) for the (1000) surface. Its surface energy (5.3 eV/nm$^2$) however is almost the double of that of the (0001) surface. The low work function does not (significantly) depend on the interslab separation and we argued that this material is not as reactive as calcium. The lower work function and reduced reactivity make Ca$_2$N a promising metal to replace Ca in cathodes of polymer light emitting diodes.

The authors wish to thank Dr. C.M. Fang (TUE) and Prof. Dr. R. Coehoorn (Philips) for helpful discussions. This work was part of the research programme of the Stichting voor Fundamenteel Onderzoek der Materie (FOM) with financial support from the Nederlandse Organisatie voor Wetenschappelijk Onderzoek (NWO).

\end{multicols}

\begin{references}
\bibitem[24]{per} Periodicity is required in all directions. This has the advantage that there is only one surface and no possible influences from other ones as is the case in experiments.
\bibitem[27]{surf} \E{S} is defined as the energy of the slab minus the energy of the equivalent bulk, divided by two times the surface area because a slab has two surfaces.
\end{references}
\end{document}